# The Multiple Merger Assembly of a Hyper-luminous Obscured Quasar at redshift 4.6

T. Díaz-Santos,[1,*] R. J. Assef,[1] A. W. Blain,[2] M. Aravena,[1] D. Stern,[3] C.-W. Tsai,[4] P. Eisenhardt,[3] J. Wu,[5] H. D. Jun,[6] K. Dibert,[7] H. Inami,[8] G. Lansbury,[9] F. Leclercq[8]

[1]Núcleo de Astronomía, Facultad de Ingeniería y Ciencias. Universidad Diego Portales, Ejército Libertador 441, Santiago, 8320000, Chile.

[2]University of Leicester, Physics and Astronomy, University Road, Leicester LE1 7RH, UK.

[3]Jet Propulsion Laboratory, California Institute of Technology, 4800 Oak Grove Dr., Pasadena, CA 91109, USA.

[4]Department of Physics and Astronomy, University of California, Los Angeles, CA 90095, USA

[5]National Astronomical Observatories, Chinese Academy of Sciences, 20A Datun Road, Chaoyang District, Beijing 100012, China.

[6]School of Physics, Korea Institute for Advanced Study, 85 Hoegiro, Dongdaemun-gu, Seoul 02455, Korea.

[7]Department of Physics, Massachusetts Institute of Technology, 77 Massachusetts Avenue, Cambridge, MA, 02139, USA.

[8]Univ Lyon, Univ Lyon1, École Normale Supérieure de Lyon, Centre National de la Recherche Scientifique, Centre de Recherche Astrophysique de Lyon UMR5574, 69230, Saint-Genis-Laval, France.

[9]Institute of Astronomy, University of Cambridge, Madingley Road, Cambridge, CB3 0HA, UK

*Correspondence to: tanio.diaz@mail.udp.cl

## Abstract

Galaxy mergers and gas accretion from the cosmic web drove the growth of galaxies and their central black holes at early epochs. We report spectroscopic imaging of a multiple merger event in the most luminous known galaxy, WISE J224607.56−052634.9 (W2246−0526), a dust-obscured quasar at redshift 4.6, 1.3 Gyr after the Big Bang. Far-infrared dust continuum observations show three galaxy companions around W2246−0526 with disturbed morphologies, connected by streams of dust likely produced by the dynamical interaction. The detection of tidal dusty bridges shows that W2246−0526 is accreting its neighbors, suggesting merger activity may be a dominant mechanism through which the most luminous galaxies simultaneously obscure and feed their central supermassive black holes.

Structure formation in the early Universe proceeds by the hierarchical assembly of dark matter haloes and the galaxies they host, with the densest structures collapsing first (*1*). During the periods of most intense accretion and growth, galaxies and their central super-massive black hole (SMBH) are expected to be obscured by interstellar gas and dust (*2*). The obscuring material absorbs the ultraviolet and optical light, from stars and the active galactic nucleus (AGN) powered by the SMBH, and re-emits it in the infrared (*2*).

The Wide-field Infrared Survey Explorer (WISE) (*3*) space telescope studied the formation and evolution of galaxies in the high-redshift Universe. A population of hyper-luminous obscured quasars were found at redshifts ≳2, with bolometric luminosities $L_{bol} \gtrsim 10^{13}$ $L_\odot$ (*4, 5*), where $L_{bol}$ is the luminosity integrated across the entire electro-magnetic spectrum, and $L_\odot$ is the luminosity of the Sun. Known as hot, dust obscured galaxies (Hot DOGs), these systems are mainly powered by accretion onto their central SMBH, which may be radiating close to the limit allowed by its own gravity (*6*). The dominant mechanism supplying the material necessary to sustain such high luminosities remains unknown. Rapid growth of galaxies and SMBHs can be accomplished via galaxy mergers, which would effectively funnel low angular momentum gas into the central AGN (*7, 8*). If Hot DOGs form in over-dense environments, merger-driven instabilities could deliver large amounts of gas and dust to the galaxy. However, there is only indirect, statistical evidence that this is the case. Combined observations of 10 Hot DOGs at 850 μm reveal more than twice as many sources within 1.5 arcminutes as in random fields (*9*), and shallow near-infrared (NIR) images show that the number density of red sources within 1 arcminute of Hot DOGs is, on average, higher than field galaxies (*10*). In sub-millimeter observations of 10 Hot DOGs, the cumulative number counts of companion sources also support dense environments around Hot DOGs (*11*). However, the morphological evidence for dynamical interactions in individual systems remains elusive, and observations obtained with the Hubble Space Telescope of a sample of Hot DOGs at redshift z ~ 2 have yielded ambiguous results (*12, 13*).

With $L_{bol}$ = 3.5 × 10$^{14}$ $L_\odot$ (*14; 15*), the Hot DOG WISE J224607.56−052634.9 (hereafter W2246−0526) is the most luminous galaxy known. Previous observations with the Atacama Large Millimeter/Sub-millimeter Array (ALMA) of the ionized carbon ([C II]) emission line at 158 μm (*16*), have shown W2246−0526 is located at a redshift of 4.601, which is equivalent to ~ 1.3 Gyr after the Big Bang, assuming standard cosmological parameters (*15*). The line profile shows a uniform, large velocity dispersion, with a full-width at half-maximum (FWHM) of ~ 500–600 km s$^{-1}$ across the whole galaxy where emission is detected (~ 2.5 kilo parsec (kpc)) (*16*). This suggests a highly turbulent interstellar medium (ISM), likely resulting from the energy and momentum injected by the central SMBH into the surrounding gas.

The [C II] observations had also shown two nearby companion galaxies to W2246−0526 (*16*). We detected a third companion in a blind search for emission line sources in our data cube, which we confirmed via its Lyman-α emission line in an optical spectrum obtained with the Keck telescope (*15*). The star formation rates (SFR) of the companions based on their [C II] luminosities are at least 7–27 M$_\odot$ yr$^{-1}$ (*15*). Other recent studies have also identified companion galaxies close to un-obscured high redshift quasars. ALMA [C II] observations of luminous quasars at z ~ 4.8 and > 6 (*17, 18*), show that a large fraction of them are accompanied by actively star-forming galaxies at projected distances < 100 kpc and within radial velocities ~< 600 km s$^{-1}$. However, there were no direct morphological signatures showing dynamical interaction between the companion galaxies and central source.

We present deep ALMA observations of the dust continuum emission at rest frame 212 μm in W2246−0526 at an angular resolution of ~0.5 arcseconds (″), equivalent to ~3.3 kpc at that distance *(15)*. The dust continuum map shown in Figure 1 reveals bridges of material connecting the central galaxy to the companions, which we denote C1, C2 and C3. The detection of dust indicates that the gas associated to these structures has been already enriched with elements heavier than hydrogen or helium, and is thus not primordial. C2 has a stream of dust extending like a tidal tail all the way to W2246–0526, over at least 35 kpc (Fig. S2). One of the densest regions in this structure, denoted K1 and located ~1.5″ northwest of C2 (Figure 1), has a counterpart identified in rest-frame near ultraviolet (UV) emission (Figure 2). The UV image was obtained with the Hubble Space Telescope (HST) using the near-IR F160W filter, and was previously analyzed in *(16)*. The detection of the UV counterpart suggests that at least a fraction of the dust in the tidal tail could be heated by in-situ star formation *(15)*. Low-surface-brightness dust emission south of C3 is coincident with strong UV emission seen in the HST image as well (denoted U1, Figure 2), although this source could be at a different redshift since it is not detected in [C II] line emission. Two more sources with unknown redshifts (U2, U3) are also identified 6″ to the northwest and 8.5″ to the southwest of W2246−0526, respectively, in the HST image.

The detection of streams of dust emission on such large physical scales suggests that W2246–0526 is in the process of accreting its neighbors—or a least stripping a large fraction of their gas—and provides evidence that (a) merger activity is taking place, and (b) the entire system may be undergoing a morphological transformation. Gas and dust accretion triggered via galaxy mergers can provide strong, yet probably intermittent, influx of material towards the nuclei of high redshift hyper-luminous galaxies, simultaneously feeding and obscuring their SMBHs.

We performed additional observations with the Karl G. Jansky Very Large Array (VLA) of the J=2→1 transition line of carbon monoxide (CO) in W2246−0526 *(15)*. The luminosity of low-J transitions of the CO molecule is regularly used as a proxy for the cold molecular gas content of galaxies. Figure 3 compares the CO(2→1) line map with contours of the 212 μm dust continuum. The CO(2→1) emission is marginally resolved (the beam FWHM is 2.47″ × 2.01″, equivalent to ~ 16 × 13 kpc), with tentative low-surface-brightness regions extending toward the companion C3 and in the direction of the tidal tail. The FWHM of the CO line in the central beam is ~ 600 km s$^{-1}$, similar to that of the [C II] line (*16,* Figure S3), suggesting that the cold molecular gas phase of the ISM traced by CO in W2246–0526 is also very turbulent, and probably affected by the strong feedback from the AGN on scales of at least a few kpc.

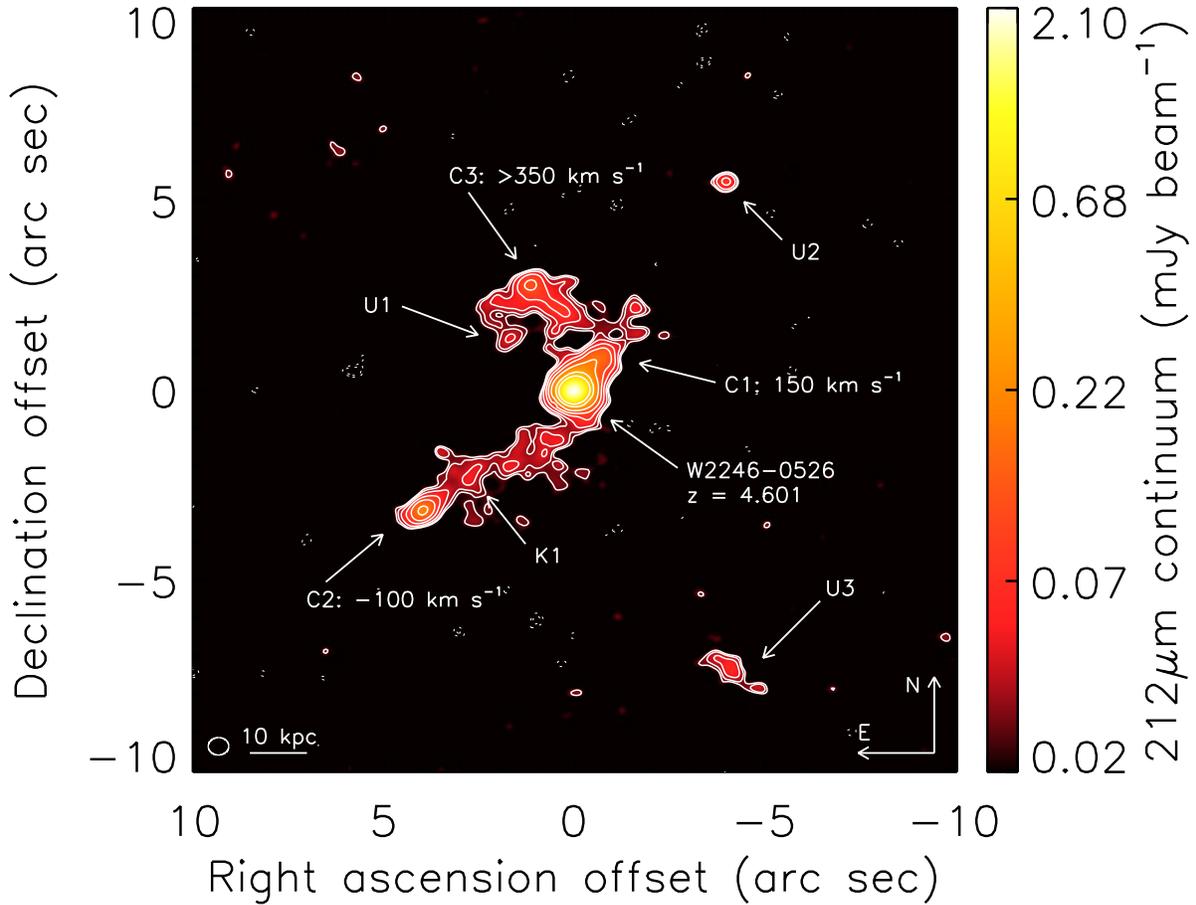

Figure 1. **ALMA 212μm dust continuum emission map of the W2246−0526 merger system.** The color bar shows the flux density on a logarithmic scale, in units of milli-jansky (1 mJy = $10^{-26}$ erg s$^{-1}$ cm$^{-2}$ Hz$^{-1}$). The north is up and the east is to the left. The angular resolution (beam size FWHM) of the observations is 0.54″ × 0.46″, or ~ 3.6 × 3.1 kpc at the redshift of W2246−0526, and it is shown by the ellipse at the bottom left corner. The offsets in the right ascension and declination axes are given in ″ relative to the center of W2246−0526, whose coordinates are: 22h 46m 07.55s, −05° 26′ 35.0″. The relative velocities of three companion galaxies (labeled as C1, C2, C3) and the redshift of W2246−0526 are measured via the [C II] emission line *(16)*, and suggest that W2246−0526 and its companions are gravitationally bound. A stream of dusty material resembling a tidal tail connects W2246−0526 with C2, and bridges join the central galaxy with C1 and C3. Three sources with unknown redshifts and the knot ~ 1.5″ northwest of C2, are labeled as U1, U2, U3 and K1, respectively. Solid contours represent levels of [2.5, 3, 4, 6, 9, 15, 30, 50] × σ, where σ is the measured root mean square (r.m.s.) of the background. Dotted contours indicate [-2.5, -3] × σ negative flux. An equivalent map with lower-significance contours is shown in Figure S2.

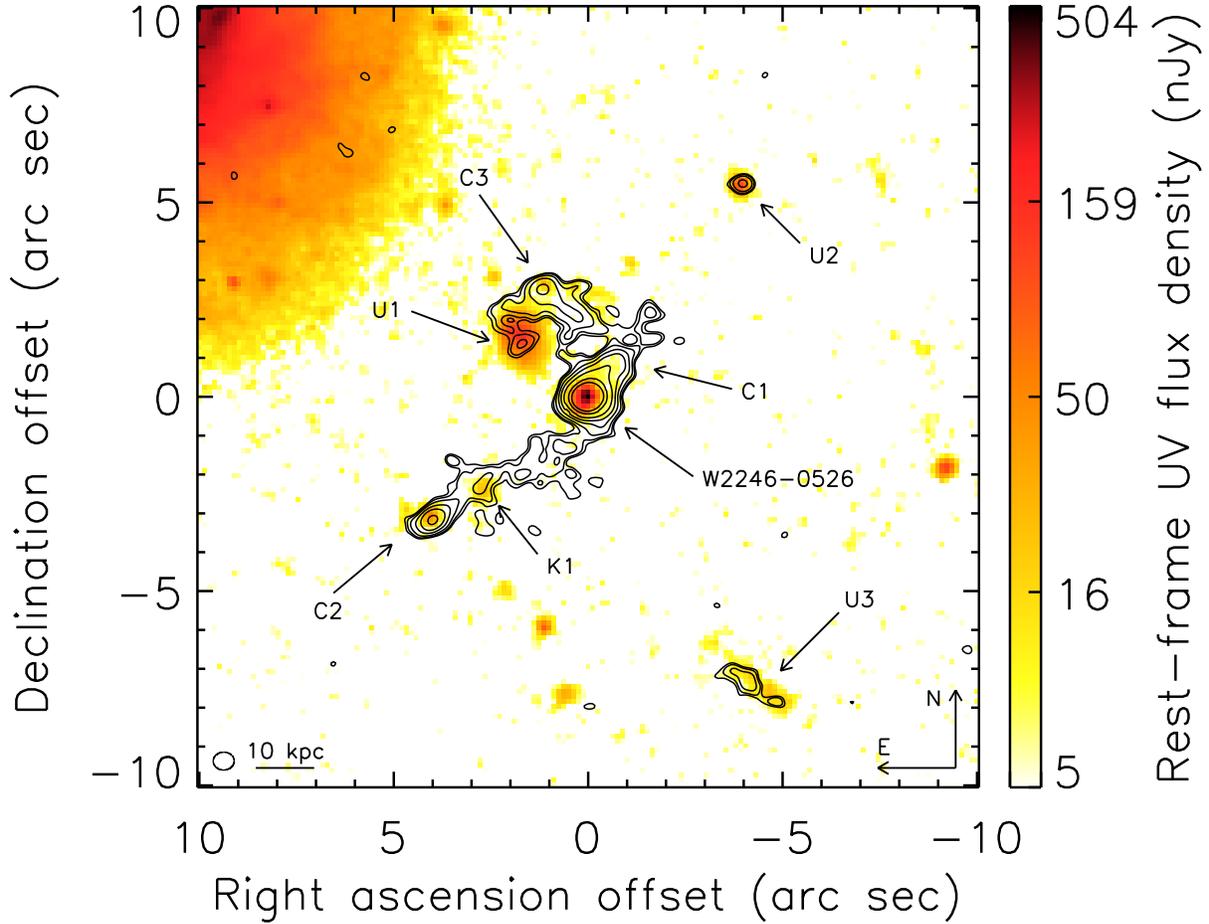

Figure 2. **Rest-frame near UV (~ 2860Å) continuum image of W2246−0526 with overlaid contours of the 212 μm dust-continuum map.** The color bar shows the near-UV flux density on a logarithmic scale. The emission at the top left corner is from a low-redshift foreground galaxy. The companion galaxies C1, C2 and C3 are detected in both the dust continuum and near-UV emission, as are the sources with unknown redshifts (U1, U2 and U3) and the tidal tail knot (K1).

The dust mass in W2246−0526 alone (within the central 1″ ~ 7 kpc) is in the range 5.6–17 × $10^8$ $M_\odot$ (assuming a dust temperature 100–50 K) (*15*), where $M_\odot$ is the mass of the Sun. This is similar to the total dust mass content of dusty ultra-luminous infrared galaxies (ULIRGs) in the nearby Universe, which span a range between ~ $10^8$ and $10^9$ $M_\odot$ (*19*) (although W2246−0526 is 100 times more luminous). The rest of the system, including the companion galaxies C1, C2, and C3, and the extended emission, contains at least as much dust as W2246−0526 alone (Table S1). The three companions contribute ~ 25% of the dust mass outside W2246−0526, and ~ 13% of the entire merger system. U2 and U3 are not included in this calculation because they may not be part of the system. The tidal tail contains almost as much dust as the sum of all three companion galaxies. Assuming that the dust and gas are well mixed, and a standard gas-to-dust ratio ($\delta_{GDR}$) typical of local, solar-metallicity galaxies (*20*), $\delta_{GDR} = 100$, W2246−0526 harbors a total gas mass ($M_{gas}$) reservoir of ~ 0.6–1.7 × $10^{11}$ $M_\odot$, with the entire system containing $M_{gas}$ ~ 1.2–3.6 ×

$10^{11}$ M$_\odot$. The total molecular gas mass estimated from the CO(2→1) line is 1.5 (± 0.8) × $10^{11}$ M$_\odot$ *(15)*, in agreement with the estimate from the dust.

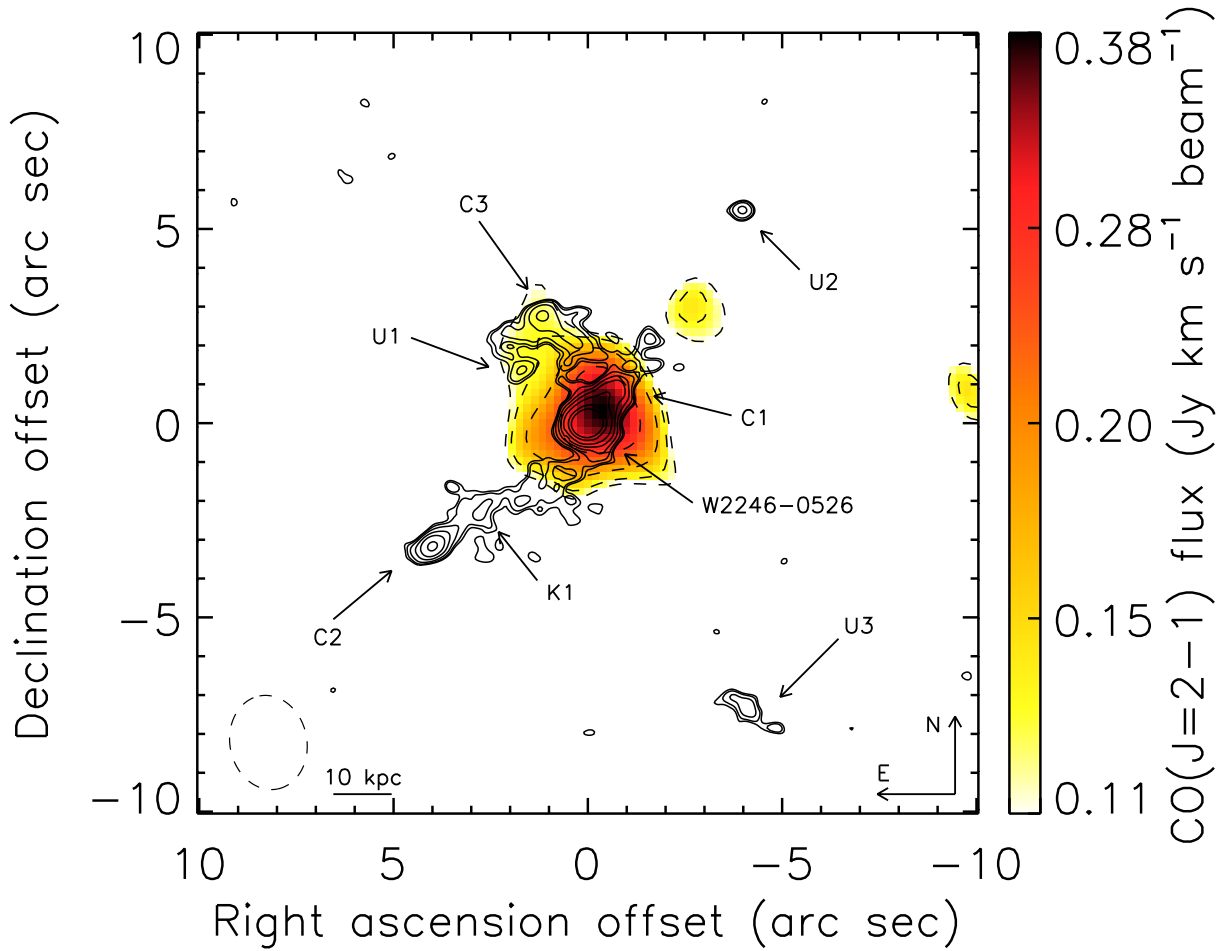

Figure 3. **CO(2→1) emission line map of W2246−0526 with overlaid contours of the 212 μm dust-continuum map.** The color bar shows the line flux per beam on a logarithmic scale. The angular resolution (beam size FWHM) of the observations is 2.47″ × 2.01″, or ~ 16 × 13 kpc at the redshift of W2246−0526, and it is illustrated by the dashed ellipse. The dashed contours represent CO levels of [2.5, 3, 4, 6, 9] × σ, where σ is the measured r.m.s. of the background. The solid contours are the same as in Figure 1, showing the ALMA 212 μm dust-continuum map. The CO(2→1) emission is only marginally resolved, slightly extending toward the companion C3, northeast of W2246−0526, and southeast in the direction of the tidal tail.

Based on the [C II] kinematics, we calculate the dynamical mass ($M_{dyn}$) of W2246−0526 to be ~ 0.8 (± 0.4) × $10^{11}$ M$_\odot$, which is within a factor of ~2 of the baryonic mass of the galaxy *(15)*, similar to observations of some compact galaxies at z ~ 2 *(21)*. The dynamical mass favors the lower bound of the dust-based gas estimate *(15)*, as expected if most of the dust within the central core of W2246−0526 (within a few kpc) is being heated to temperatures > 100 K, due to its closer proximity to the central AGN.

The $M_{gas}$ and stellar mass ($M_\star$) in the W2246−0526 merger system imply a baryonic gas fraction $f_{gas}$ ~ 0.3–0.6 (where $f_{gas} = M_{gas}/(M_\star+M_{gas})$), which is lower than the value expected for most galaxies at similar redshifts (*22*). The SFR within the central ~ 4″ is estimated to be ~ 560 M$_\odot$ yr$^{-1}$, with a factor of two uncertainty (*15*). This translates into a specific SFR (SSFR, the star formation rate divided by the stellar mass) of ~ 2.2 Gyr$^{-1}$, equivalent to a mass doubling-time of ~ 450 Myr, both with an uncertainty of a factor of three. The derived SSFR is only slightly lower than that of main-sequence galaxies at redshifts 3.5–5 (*22, 23*). However, the gas depletion time scale is only ~125–700 Myr (*15*), which is between main-sequence and starburst galaxies at approximately the same redshift (*22*). The free-fall time of the gas in the system is ~ 100–170 Myr (*15*), at least an order of magnitude larger than the active period of hyper-luminous quasars (*24*), suggesting that the Hot DOG phase is shorter than the dynamical time scale of the merger. Based on the $M_{gas}$ of the tidal tail and companion galaxies, the average accretion rate of gas towards the center of W2246−0526 could be as high as $dM_{gas}/dt$ ~ 550–900 M$_\odot$yr$^{-1}$ (*15*), similar to the estimated SFR of the underlying galaxy.

We interpret these results as showing that merger-driven accretion of neighbor galaxies can be a catalytic mechanism that simultaneously i) obscures the central SMBH in W2246−0526 under large columns of dust and gas, and ii) provides the intermittent, large-scale influx of material needed to generate its extreme luminosity, and maintain star formation in the host galaxy, which would otherwise quickly deplete its gas reservoir. The energetic AGN feedback resulting from this accretion is likely responsible for maintaining the turbulence of the gas at the center of W2246−0526. Slow, nearly-isotropic ISM outflows on scales of a few kpc can coexist with the accretion of material stripped from in-falling galaxies at larger scales (even if the companions themselves may only fly by), which can be funneled efficiently into the central AGN through collimated, filamentary structures (*25, 26, 27*). If W2246−0526 is representative of the Hot DOG population, our results suggest that hyper-luminous obscured quasars may be interacting systems, the result of ongoing merger-driven peaks of SMBH accretion and massive galaxy assembly in the early Universe.

**Acknowledgments:**

We thank Adam Stanford and Mislav Baloković for helping to obtain optical spectra for W2246−0526 on November 2010 and October 2013, respectively. We thank Jorge González López for helpful suggestions regarding the cleaning algorithms of CASA.

**Funding**:

T.D-S. acknowledges support from ALMA-CONYCIT project 31130005 and FONDECYT project 1151239. R.J.A. acknowledges support from FONDECYT 1151408. The work of C-W.T., J.W., P.E., and D.S. was carried out at the Jet Propulsion Laboratory, California Institute of Technology, under a contract with NASA, and supported by grant ADAP13-0092. M.A. acknowledges partial support from FONDECYT through grant 1140099. J.W. acknowledges support from MSTC through grant 2016YFA0400702 and NSFC 11673029. This research was supported by the Basic Science Research Program through the National Research Foundation of Korea (NRF) funded by the Ministry of Education (NRF-2017R1A6A3A04005158).

**Author contributions:**

T.D-S. has lead the overall project. R.A. and A.W.B. have contributed with their expertise to the interpretation of the results. M.A. has helped with the processing of the ALMA and VLA data. D.S. led the spectroscopic identification campaign of Hot DOGs. C.-W.T., P.E. and J.W. have provided their expertise in the discussion of the results. K.D., H.I, G.L. and F.L were part of the team that acquired the optical spectra of W2246−0526.

**Data and materials availability:**

**Supplementary Materials:**

Materials and Methods

Figures S1-S4

Tables S1

**Materials and Methods**

Keck/LRIS spectrum

We observed W2246−0526 multiple times with the dual-beam Low Resolution Imaging Spectrometer (LRIS) *(28)* on the Keck I telescope. All observations used the 400 lines mm$^{-1}$ grating on the red arm of the spectrograph (blazed at 8500 Å; resolving power R ~ 750), with the 5600 Å dichroic, and the 1.5″ wide long slit. W2246−0526 is not detected shortward of the dichroic wavelength, and so we do not discuss the data from the blue arm further.

Our first spectrum, obtained on Universal Time (UT) 2010 November 8, entailed two 600-s exposures at a position angle (P.A.) = 49.9° (east of north). A relatively narrow (~ 600 km s$^{-1}$), asymmetric Lyman-α (Lyα) emission line was detected (see Figure S1), with some faint Lyα nebulosity. Combined with a modest continuum break across the Lyα emission line and a low signal-to-noise ratio detection of a matching broad (~ 4000 km s$^{-1}$) C IV emission line, these data yielded a redshift of z = 4.6.

We obtained a deeper follow-up spectrum on UT 2013 October 4 consisting of three 1200-s exposures at a P.A. = 38° in photometric conditions. The deeper spectrum confirmed the features observed in the 2010 data, with higher signal-to-noise ratio (Figure S1). A Lyα redshift of 4.601, based on a Gaussian fit to the peak of the emission line, matches the ALMA-derived [C II] redshift of 4.601 ± 0.001. The C IV line detected by LRIS yields a lower redshift of $z_{CIV}$ = 4.548 (blue-shifted ~ 2800 km s$^{-1}$ from the Lyα line) based on the peak of the highly asymmetric, broad feature. This is a larger offset than typically seen in lower redshift quasars *(29)*, and more typical of the high redshift (z > 6) quasar population *(30)*. Such large blue-shifts are suggestive of high radiative efficiencies and very strong winds *(31)*.

We obtained a third spectrum of W2246−0526 on UT 2015 December 12 (Figure S1 inset), consisting of two 1200-s exposures and a single 462-s exposure at P.A. = 20° **(see Figure S2)**. Though the night was clear, there were 40-50 mph winds, which produced poor seeing due to wind-shake (~ 1.5″). These data reproduce the features seen previously, namely narrow Lyα emission and blue-shifted, broad C IV emission. The 2D spectrum also showed a faint, slightly red-shifted Lyα emission feature extending ~ 3″ to the northeast of W2246−0526, spatially coincident with the companion galaxy C3 (see Figure S2), further supporting its detection and the derived redshift estimated from the [C II] ALMA data. Unfortunately, the low angular resolution of the optical spectrum does not allow the identification of Lyα emission lying between W2246−0526 and C3.

ALMA 212μm Dust Continuum Data

The ALMA observations of W2246−0526 in the ~ 212 μm dust continuum emission were obtained in Band 6 (211–275 GHz) during 3 execution blocks, two on 2016 June 20 and one on 2016 July 13. The on-source integration time was 49 min per block, for a total of 2.5 hours. The number of antennas used in each run was 38, 41 and 39, respectively. The minimum and maximum baseline lengths were 15.1 and 704.1 m. The sources Pallas and QSO J2148+0657 were used for amplitude calibration, the source QSO J2232+1143 was used for pointing and band-pass calibrations, and the source PMN J2243−0609 was used for phase calibration.

We used the Common Astronomy Software Application (CASA; v.5.1.1) *(32)* to process and clean the ALMA products. All the execution blocks of the source were concatenated in a single measurement set for the analysis. The cleaning algorithm was run using the task *tclean* in parallelized process, with a *briggs* weighting scheme, a *hogbom* deconvolver, and a *robust* parameter set to = 2 (similar to natural weighting) for the *u-v* visibility plane. A single circular aperture of radius = 1″ was used to mask the core of W2246−0526, cleaning each window down to a depth of 2 times the r.m.s.. The angular size of the restoring beam ranges from 0.52″ × 0.44″ in the spectral window (SPW) 0 (central observed frequency $\nu_{obs}$ ~ 260.929 GHz; rest wavelength $\lambda_{rest}$ ~ 205 µm) to 0.56″ × 0.48″ in the SPW 3 (central $\lambda_{rest}$ ~ 219 µm), with an average of 0.54″ × 0.46″ at an effective $\lambda_{rest}$ ~ 212 µm, or ~ 3.6 × 3.1 kpc at the redshift of W2246−0526. The P.A. of the beam is ~ 80°. The average r.m.s. of the data cube is ~ 125 µJy beam$^{-1}$ channel$^{-1}$, measured in 48 km s$^{-1}$ channels (averaged over 5 original channels). To create the moment-0 map shown in Figure 1 and Figure 2S, we collapsed the dust continuum emission of all four SPWs after discarding bad channels at the edges of the SPWs, reaching a r.m.s. = 11.5 µJy beam$^{-1}$.

Figure 1 shows the dust continuum map of the W2246−0526 merger system down to a lowest contour of 2.5 × σ, where σ is the r.m.s. of the background. This is a sufficiently high threshold for the map not to show a large number of regions with negative flux, implying that the fidelity of the positive emission detected is high. However, for reference, we also show in Figure S2A a map where an additional contour is plotted at 2 × σ (as well as the complementary -2 × σ contour). While noisier, this figure shows that most of the positive emission recovered at the 2 × σ level is located around the tidal tail and the system of galaxies in general.

To check the reliability of the extended emission in the dust map, we processed and analyzed each execution block independently, which were obtained at different periods of time, thus having different *u-v* coverage. We followed the same cleaning procedure used for the combined dataset described above. Albeit with a lower signal-to-noise, extended emission was detected in between W2246−0526 and the companion galaxy C2 in every execution block We have also processed and analyzed the dataset of the "check source". The check source is a bright, point-like object that is always observed at the end of an execution block, which can be used to double-check the calibration of the science target. The image of the check source does not show any trace of residual/artifact emission in the direction of the tidal tail detected in W2246−0526, which was observed with the same configuration.

We also cleaned the image of W2246−0526 using a *uv*-tapering resulting in a FWHM ~ 1″. The tidal tail is recovered with a higher significance at larger angular scales (Figure S2B).

VLA CO(2→1) Data

Observations of the redshifted $^{12}$CO J = 2–1 emission line ($\nu_{rest}$ = 230.538 GHz; $\lambda_{rest}$ ~ 1.3 mm) in the W2246−0526 system were obtained using the VLA in 5 observing runs between 2015 November 02 and 2015 December 07. The observations were taken in D-configuration with 27 antennas, using the Q-band receivers (tunable range: 40–50 GHz). At the redshift of W2246−0526 the CO(2→1) line is redshifted to 41.160 GHz.

The observations were performed with the Wideband Interferometric Digital ARchitecture.(WIDAR) correlator, with two basebands (AC and BD) of eight contiguous SPWs each. Baseband AC covered the SPWs 0–7 while baseband BD covered SPWs 8–15. Each SPW was set to have 64 channels and 2 MHz per channel resolution. Both basebands were configured to overlap, with center frequencies 41.128 GHz and 41.228 GHz, for AC and BD respectively, making an effective bandwidth of 1.120 GHz. With this setup, the SPWs 4 and 11 covered the redshifted CO(2-1) line, being tuned to 41.160 and 41.192 GHz, respectively.

The nearby quasar QSO J2229-0832 was used for gain and pointing calibration and the source 3C 048 served as flux and bandpass calibrator. The data were calibrated using the Astronomical Image Processing System (AIPS) *(33)* and CASA *(32)*. Time ranges with poor visibilities as well as edge channels where the bandpass deteriorated (at each edge) were flagged. The data were imaged using the *tclean* algorithm in CASA. All images were primary beam corrected. We used a *briggs* weighting scheme, a *hogbom* deconvolver, and cleaned down to 2σ in a circular aperture of ~15″-diameter around our target. Setting the *robust* parameter = 2 (similar to natural weighting), results in an angular resolution of 2.47″ × 2.01″ (P.A. = 10.4°). We tried other weighting schemes, but found that most of the emission was not recovered when using lower values for the robust parameter. The final cube is shown in Figure 2 and has a r.m.s. = 85 μJy beam$^{-1}$ channel$^{-1}$ measured in 60 km s$^{-1}$ channels. The continuum emission under the CO line was not detected. The r.m.s. of the collapsed cube is ~ 10 μJy beam$^{-1}$.

Figure S3 presents a comparison between the spectra of the CO(2→1) and [C II] 158 μm emission lines *(16)*. Both lines display high velocity dispersions, FWHM ~ 550 – 600 km s$^{-1}$, suggesting that both the neutral and molecular gas phases of the ISM are very turbulent, probably reflecting the energy and momentum that are being injected by the central AGN on its surrounding medium.

Cosmology

Throughout the paper, we adopt a cosmology with the following parameters: $\Omega_M = 0.28$, $\Omega_\Lambda = 0.72$ and $H_0 = 70$ km s$^{-1}$ Mpc$^{-1}$. At the redshift of W2246−0526, z = 4.601, the angular scale is 1″ = 6.68 kpc.

Star Formation Rate from the [C II] Emission Line

The [C II] 158 μm emission line has been proposed as a reliable SFR tracer in normal, Milky Way-like, star-forming galaxies as the [C II] luminosity correlates well with other SFR indicators *(34)*. However, the [C II] luminosity of IR-bright sources fall below this trend, displaying a "deficit" of line emission with respect to the total IR emission *(35-37)*, which scales linearly with the SFR in dust-obscured objects. The [C II] to IR luminosity ratio ($L_{[CII]}/L_{IR}$) in luminous infrared galaxies can be, in extreme cases, a factor of ~ 20 smaller than in normal star-forming galaxies, which show a typical value $L_{[CII]}/L_{IR} \sim 5 \times 10^{-3}$ *(36)*. Thus, if no additional information other than the IR luminosity of the source is available, a detection of the [C II] emission can only provide a lower limit to the SFR of the galaxy. We use a calibration between $L_{IR}$ and SFR *(38)*, with SFR$_{IR}$ [M$_\odot$ yr$^{-1}$] = 1.5 × 10$^{-10}$ $L_{IR}$ [L$_\odot$], based on the Starburst99 stellar evolution synthesis models *(39)*, and assuming a constant SFR history over 100 Myr, a Kroupa initial mass function (IMF) *(40)*, and that the entire Balmer continuum emitted by the starburst is absorbed and re-radiated by optically thin dust.

The [C II] luminosity of W2246−0526 measured with a 1″-diameter aperture is $L_{[CII]}$ = 6.1 × 10$^9$ L$_\odot$ *(16)*. The line is spatially resolved, with up to ~ 55% of the emission arising from an extended component (measured in the same aperture). The extended emission thus accounts for $L_{[CII]} \sim 3.4 \times 10^9$ L$_\odot$. If we assume that the extended [C II] emission is powered entirely by star formation, then using the $L_{[CII]}/L_{IR}$ ratio upper limit for normal galaxies and the SFR calibration described above, we estimate that the lower limit to the SFR of the host is $\gtrsim$ 100 M$_\odot$ yr$^{-1}$. If the star formation properties of the underlying galaxy in W2246−0526 are instead similar to local, purely star-forming ULIRGs, its SFR could be as high as ~ 1000 M$_\odot$ yr$^{-1}$. For reference, the SFR derived via SED fitting for the central ~ 4″ of W2246−0526 is ~ 560 M$_\odot$ yr$^{-1}$ (see below).

The [C II] luminosities of the companion galaxies C1, C2 and C3 are (9.1, 6.6, $\gtrsim$ 2.3) × 10$^8$ L$_\odot$, respectively. Assuming that all the IR luminosity in these galaxies is associated with star formation, we estimate SFR lower limits of $\gtrsim$ 27, 20 and 7 M$_\odot$ yr$^{-1}$, respectively, and upper limits in the range of 200–500 M$_\odot$ yr$^{-1}$, when using a $L_{[CII]}/L_{IR}$ ratio typical of ULIRGs.

Calculation of Dust and Gas Masses

To estimate the dust mass content of W2256−0526 and its neighboring galaxies, we use a modified blackbody function scaled to the continuum flux density of each source, as measured at the average rest-frame wavelength of the ALMA data cube, $\lambda_{rest} \sim 212$ μm. We further assume that the dust emission is optically thin, with an emissivity index $\beta = 1.8$ *(41)*. We adopt a dust mass opacity $\kappa_{\nu,dust}$ (850 μm) = 0.0484 m² kg⁻¹, which assumes a gas-to-dust mass conversion factor typical of local solar-metallicity galaxies, $\delta_{GDR} = 100$ *(41,42)*, consistent with the gas mass estimates based on the CO(2→1) emission line. The adopted $\kappa_{\nu,dust}$ is close to the value $\kappa_{\nu,dust}$ (800 μm) = 0.04 m² kg⁻¹ obtained by *(43)*. Under these assumptions, the dust mass is $M_{dust} = (S_{\nu,obs} \times d_L^2)/ (\kappa_{\nu,rest} \times B(T_{dust}) \times (1+z))$, where $S_{\nu,obs}$ is the measured flux density of the dust continuum at the observed frequency $\nu_{obs}$, $d_L$ is the luminosity distance to the source, $B(T_{dust})$ is the Planck function evaluated at the dust temperature, $T_{dust}$, and $z$ is the redshift of the galaxy.

This approach usually assumes that most of the dust and ISM mass in a galaxy is accounted by its coldest component *(22)*. As such, instead of using a luminosity-weighted dust temperature in the equation above, $<T_{dust}>_L$ (derived from fitting the IR spectral energy distribution (SED) of the galaxy, or the position of its far-IR emission peak), $T_{dust}$ is normally replaced by a mass-weighted temperature, $<T_{dust}>_M = 25$ K in normal star-forming galaxies, more representative of the bulk of their cold dust and ISM content. However, given the extreme luminosities and dust temperatures of Hot DOGs this assumption may not be appropriate, and could lead to overestimate their dust masses. Using a single-temperature modified blackbody model, the coldest component of the $T_{dust}$ distribution in a sample of Hot DOGs is found to be in the range ~ 60–120 K *(4)*. Thus, to calculate the dust mass in W2256−0526, we adopt a reasonable temperature range of $T_{dust} = 50–100$ K for the Rayleigh-Jeans tail emission, with the lower end providing the largest $M_{dust}$ estimations. For the companion galaxies and extended emission in general, we use the narrower range of $T_{dust} = 25–50$ K. A summary of the dust mass estimates can be found in Table S1. For a discussion of the possible heating mechanisms of the dust in the tidal tail, see below.

At the average rest-frame wavelength of ~ 212 μm, the observed flux density of the dust continuum emission within the central ~ 7 kpc (1″-diameter) of W2246−0526 (i.e., without including its neighbors) is 2.3 ± 0.1 mJy. This corresponds to a dust mass in the range $M_{dust} \sim 5.6–17 \times 10^8$ M$_\odot$ ($T_{dust} = 100–50$ K, respectively). The flux density of the extended structures and diffuse emission around W2246−0526 (including the neighbor galaxies as well as the tidal tail and bridges) is 2.6 ± 1.3 mJy, and thus has a similar estimated dust mass. The companions C1, C2 and C3 have, respectively, individual flux densities of 0.27, 0.23 and 0.16 mJy in a 1″-diameter aperture, and account for ~ 25% of the extended emission. Using a gas-to-dust mass ratio $\delta_{GDR} = 100$, the total gas reservoir in W2246−0526 alone is $0.6–1.7 \times 10^{11}$ M$_\odot$, and the entire system could contain as much as $M_{gas} = 1.2–3.6 \times 10^{11}$ M$_\odot$.

Measuring the CO(2→1) line flux in a circular aperture with a radius equivalent to the beam size of the observations (~ 4.4″-diameter, which includes W2246−0526 and C1, see Figure 1) and following *(44)*, the luminosity of the line is $L_{CO(2\to1)} = 3.3$ (±0.7) $\times 10^7$ L$_\odot$. Assuming that the CO is in local thermal equilibrium up to the J = 2–1 transition and using a $M_{gas}$-$L_{CO(1\to0)}$ ratio, $\alpha_{CO}$, typical of local normal, Milky-Way-like galaxies ($\alpha_{CO,MW} = 4.6$ M$_\odot$ [K km s⁻¹ pc²]⁻¹ *(41)*, which does not include the factor of 1.36 to account for the gas mass in helium and heavier elements), we obtain a gas mass of $M_{gas} \sim 3.9 \times 10^{11}$ $M_\odot$. If an $\alpha_{CO}$ conversion factor typical of nearby ULIRGs is used ($\alpha_{CO,ULIRG} = 0.8$ $M_\odot$ [K km s$^{-1}$ pc$^2$]$^{-1}$) (45), then $M_{gas} \sim 0.7 \times 10^{11}$ $M_\odot$. The estimate based on the 212 μm dust continuum flux density (3.9 mJy) measured in the same aperture is in the range $\sim 1.0 - 2.8 \times 10^{11}$ $M_\odot$ (using $\delta_{GDR} = 100$), which slightly favors an $\alpha_{CO}$ conversion factor closer to the ULIRG regime. The stellar mass of the galaxy measured in similar aperture (r ~ 2″) is $2.5 \times 10^{11}$ $M_\odot$ (see below). Thus, the gas depletion time scale would be 125–700 Myr, depending on the $\alpha_{CO}$ employed.

The total gas mass measured in a 11″-diameter (~ 73 kpc) aperture that includes the bulk of the CO(2→1) emission (i.e., the entire merger system: W2246−0526 and C1, as well as C2, C3, tidal tail and bridges) is $M_{gas} \sim 1.5 \times 10^{11}$ $M_\odot$, using $\alpha_{CO,ULIRG}$. This is in agreement with the range obtained from the dust-based measurement ($M_{gas} = 1.2–3.6 \times 10^{11}$ $M_\odot$). An $\alpha_{CO,MW}$ would yield a $M_{gas} \sim 8.6 \times 10^{11}$ $M_\odot$.

Possible Heating Mechanisms for the Dust in the Tidal Tail

Physical mechanisms that could heat the dust in the tidal tail connecting W2246−0526 and the companion galaxy C2 are: (a) direct illumination from the AGN, and (b) in-situ star formation.

Scenario (a) assumes that dust in the tidal tail may be heated by the central AGN. Radiative transfer models can reproduce the spectral energy distribution of AGNs using spherical geometries where dust is smoothly distributed in a cloud that can be as large as ~ 15 kpc, surrounding the central energy source (46). The equilibrium temperature of dust heated by a central source ($T_{dust}$) can be calculated as a function of its luminosity and distance (47). For a bolometric luminosity equal to that of W2246−0526, dust can be heated up to $T_{dust} = 25$ K at distances up to ~ 7.5 and 100 kpc, for optically thick and thin material, respectively. That is, the central energy source is more likely to heat dust to a high temperature at distances > 1″ (see Figure 1) if the obscuring medium around W2246−0526 is effectively optically thin, such that the UV light is not quickly reprocessed by intervening material. This suggests that the dust geometry surrounding W2246−0526 would need to be clumpy, because virtually all the UV continuum is fully reprocessed (optically thick) along our line of sight, but the direction from the AGN towards the tidal tail would need to be effectively free of dust (mostly optically thin). If this were the case, we should see a gradient of temperatures along the tidal tail, which extends for ~35 kpc, related to the distance at which the AGN can heat the dust at a certain temperature. While the dust column density could compensate for the temperature gradient such that we would observe a uniform continuum flux density along the tail, such a scenario is rather complex.

Scenario (b) is that the dust in the tidal tail may be heated by in-situ star formation. This is disfavored by the fact that, aside from a faint detection of the knot K1, the dust emission connecting C2 and W2246−0526 as well as the bridge to C3 does not have associated UV emission detected in the HST image (Figure 2). However, the non-detection of UV continuum may simply be caused by dust obscuration (e.g., *48*), or the the shallow depth of the HST image, which probes only down to a sensitivity of SFR ~ 6 $M_\odot$ yr$^{-1}$.

Alternatively, the dust could be cooling, without being exposed to any source of heating. However, this case is highly unlikely since dust is a very efficient coolant. The cooling rate of dust at $T_{dust}$ > 50 K is extremely short ($\Lambda_{dust} \propto n\ T_{dust}^{4+\beta}$, where n is the number density of hydrogen atoms), reaching equilibrium with the cosmic microwave background (CMB) within years, thus making it effectively undetectable. In addition, while cooling, any gas clump in the tidal tail would undergo gravitational collapse. The free-fall time of a parsec-sized clump with a mass of $10^6$ $M_\odot$ is a few kyr, thus suggesting that star-formation should have already started to take place in the gas stream (if the shear is not too strong). We therefore favor scenario (b).

Estimation of the Global Stellar Mass and Star Formation Rate of the Host Galaxy Through SED Fitting

We have constructed the optical to far-IR spectral energy distribution (SED) of W2246−0526 (Figure S4) using photometry from the following telescopes: HST Wide Field Camera 3 at 1.6 μm with the F160W filter (*16*), Palomar200 at 2.2 μm with the Ks filter (*10*), Spitzer at 3.6 and 4.5 μm with the IRAC1 and IRAC2 filters (*49*), WISE at 12 and 22 μm with the W3 and W4 filters (*3*), Herschel at 250, 350 and 500 μm with the Spectral and Photometric Imaging Receiver (SPIRE) (*14*), ALMA at 865 μm and 1.19 mm in Bands 7 and 6 (*16*, and see above), and VLA at 7.28 mm in Q-band (see above). In all cases we collected the photometry with the smallest aperture available to avoid including emission from the neighbor galaxies. While the angular resolution of HST and ALMA allows us to easily separate all companions, the Spitzer, WISE and Herschel angular resolutions are coarser, and thus the fluxes could be contaminated. We use the AllWISE catalog to investigate the mid-IR WISE photometry, which is extracted via a point-spread-function (PSF) fitting procedure that also provides the goodness of fit of the model. In all WISE bands W2246−0526 is unresolved (reduced $\chi^2 \leq 1$) and no deblending was needed. Therefore, any contamination of the extended companions to the mid-IR fluxes is negligible. Regarding the near-IR bands, the Spitzer photometry was obtained with a 6″-diameter circular aperture using Sextractor (*49*). While Sextractor performs a deblending process in crowded fields, extended emission close to C3 could contribute up to 30% of the total flux in IRAC1 and 15% in IRAC2. If we were to subtract these contributions from the Spitzer photometry, the estimated stellar mass ($M_\star$) of W2246−0526 (see below) would be reduced by ~20%. Thus, even in this worst-case scenario, the change in $M_\star$ would be well within the error of a factor of two quoted below. The Herschel photometry is not used for any fitting, just to guide the eye and compare them with the extrapolation to shorter wavelengths of the fit to the ALMA data, so we do not investigate contamination of the Herschel photometry.

To estimate the stellar mass of W2246−0526 we follow the same approach as *(10)*, who modeled the optical to mid-IR SEDs of 96 Hot DOGs using the algorithm and SED templates for AGN and galaxies from *(50)*. Briefly, these are four empirical SED templates that span the wavelength range from 0.03 to 30μm. An object is modeled as a non-negative combination of an old stellar population (E template), an intermediate stellar population (Sbc template), a young stellar population (Im template) and an AGN. We also fitted a reddening component to the AGN template, with a dust extinction law that is a combination of a Small Magellanic Cloud (SMC) reddening law at shorter wavelengths and a Milky Way reddening law at longer wavelengths. Intergalactic medium absorption is also considered. We do not include an extinction component for the host galaxy because the templates already include a small amount of intrinsic reddening *(50)* and the model does not require further obscuration. We discuss below the effects of unaccounted dust obscuration for the SED model of W2246–0526.

Fig S4 shows the best-fitting SED model for W2246−0526. The host galaxy dominates the rest-frame optical wavelengths while the AGN dominates at $\lambda_{rest} \gtrsim 1$ μm. We do not expect the optical spectrum of W2246−0526 to be contaminated by direct light from the AGN because i) the HST image shows that the emission from the core is resolved and ii) the considerable dust obscuration expected towards the central engine. Nevertheless, at ~ 3000 Å there could be a contribution from scattered AGN light, as has been seen in a few Hot DOGS *(51)*, which would result in a lower stellar mass. Figure S4 shows that the observed-frame Ks-band flux density is below the best-fitting SED model, and may indicate some obscuration toward the host galaxy. However, the IRAC bands are well fitted by the model, suggesting the host-galaxy obscuration is not enough to distort the redder parts of the SED, particularly the rest-frame near-IR. If we disregard the HST F160W band and allow for host reddening, then we find that the best-fitting SED still uses the same templates, and only requires a color excess E(B-V) = 0.05 for the host. The stellar mass estimate presented below would be only 5% lower, however, which is negligible. We therefore remove the AGN contribution from the best-fitting SED model and estimate the stellar mass using the rest-frame H-band luminosity of the host galaxy template. To do this, we follow *(10)* and derive the mass-to-light ratio of the host using the code *ezgal (52)* and the Galaxev stellar population SED models *(53)*, which include a contribution from thermally pulsing-asymptotic giant branch (TP-AGB) stars, with a metallicity $Z = 0.008$ ($\equiv 0.4\ Z_{\odot}$, where $Z_{\odot}$ is the metallicity of the Sun) and a Chabrier IMF *(54)*.

We consider exponentially declining SFR histories with different decaying timescales, $\tau$. If we assume that the starburst started 100 Myr ago (i.e., at a redshift $z_f = 4.9$) and a $\tau = 1$ Gyr (similar to a constant SFR history), we estimate a stellar mass of $M_\star = 1.7 \times 10^{11}$ $M_\odot$. A shorter decaying timescale of $\tau = 100$ Myr would increase the estimated stellar mass by only 12% to $M_\star = 1.9 \times 10^{11}$ $M_\odot$. If we instead assume that the starburst started 500 Myr ago (i.e., $z_f = 6.7$) and a $\tau = 1$ Gyr, the derived stellar mass is $M_\star = 3.4 \times 10^{11}$ $M_\odot$. We do not consider estimates for shorter $\tau$, as the current SFR would be lower than a few hundred $M_\odot$ yr$^{-1}$, in disagreement with our measurements based on the IR luminosity (see below). Exponentially increasing SFR histories would yield smaller stellar masses than exponentially declining or nearly constant ones.

ALMA observations showed that the dust continuum flux density at rest-frame ~ 158 μm is 7.4 mJy *(16)*. At least 72% of the flux is accounted for by a central point source, with the remaining 28% being in an extended component, and hence likely powered by star-formation activity *(16)*. If we assume that the SED of the host galaxy is well described by a starburst similar to that in M82 *(55)*, the 8–1000μm luminosity would correspond to a SFR = 560 $M_\odot$ yr$^{-1}$ *(56)*. If we instead assume the host galaxy SED corresponds to the Sd spiral template of *(55)*, the implied SFR would be 190 $M_\odot$ yr$^{-1}$, although the compactness of the host galaxy and its turbulent nature *(16)* makes it very unlikely that there is no ongoing starburst. A SFR = 560 $M_\odot$ yr$^{-1}$ is in rough agreement with all the SFR histories used above to estimate the stellar mass of the galaxy, which predict current SFRs ~ 500 – 1500 $M_\odot$ yr$^{-1}$.

Considering that the uncertainties in stellar mass estimates are inherently high *(57)*, all values derived above are roughly consistent. Hence, for simplicity, we adopt the following properties for the integrated emission (r ~ 2″) of W2246−0526: $M_\star$ = 2.5 × 10$^{11}$ $M_\odot$ and SFR = 560 $M_\odot$ yr$^{-1}$, both with an uncertainty of a factor of 2. The SSFR of the system derived from these values is ~ 2.2 Gyr$^{-1}$, equal to a stellar mass doubling-time of ~ 450 Myr, both with an uncertainty of a factor of three.

Calculation of the Dynamical Mass and Accretion Rate

The velocity shear of the [C II] line is small *(16)* and so we can calculate, based on the [C II] kinematics, the dynamical mass of W2246−0526 by assuming that it is dispersion-supported. We obtain $M_{\rm dyn}$ = 0.8 (± 0.4) 10$^{11}$ $M_\odot$, for an intrinsic velocity width of the line FWHM = 550 (± 100) km s$^{-1}$, and a deconvolved physical diameter of the detected emission of $d$ = 2.5 (± 1) kpc. While the choice of using a dispersion-supported scenario for the calculation of $M_{\rm dyn}$ is justified by the line's large FWHM, having the system partially supported by rotation would allow for a larger $M_{\rm dyn}$ if the disk of the galaxy is inclined (face on).

The stellar mass of the system within a 4″-diameter (~ 27 kpc) is $M_\star$ ~ 2.5 × 10$^{11}$ $M_\odot$ (see above). If $M_\star$ scales with aperture in the same way as $M_{\rm dust}$, then $M_\star$ ~ 1.5 × 10$^{11}$ $M_\odot$ in W2246−0526 alone. If we use the gas mass derived from the CO(2→1) observations within the central ~4.4″ and scale it down similarly, then $M_{\rm gas}$ ~ 0.4 × 10$^{11}$ $M_\odot$ in W2246−0526 (assuming $\alpha_{\rm CO,ULIRG}$). Thus, the total (baryonic) mass $M_{\rm bar} = M_\star + M_{\rm gas}$ agrees with $M_{\rm dyn}$ within a factor of ~ 2, which is well within the uncertainties. The excess of $M_{\rm bar}$ over $M_{\rm dyn}$ puts constraints on the conversion factors used to obtain $M_{\rm gas}$ from the CO(2→1) line and from the dust continuum in W2246−0526 alone, somewhat favoring the ULIRG $\alpha_{\rm CO}$ conversion for CO(2→1) and the higher end of the $T_{\rm dust}$ range (100 K) for the dust-based estimate. This agrees with the scenario of the dust within the central core of W2246−0526 (r < few kpc) being heated by the central AGN to $T_{\rm dust}$ ~ 100 K *(4)*.

Assuming $M_{\rm bar} = M_\star + M_{\rm gas}$ ~ 4 × 10$^{11}$ $M_\odot$ for the entire merger system (see above), the free-fall time, $t_{\rm ff}$, at the position of the most distant, spectroscopically identified companion galaxy (C3, at ~35 kpc from W2246−0526) is $t_{\rm ff}$ ~ 170 Myr (~ 100 Myr if an $\alpha_{CO,MW}$ is used to derived the gas mass from the CO(2→1) line). This is equivalent to an average inflow velocity of ~ 200 km s$^{-1}$ (330 km s$^{-1}$ for $\alpha_{CO,MW}$). Assuming the most conservative value of $t_{\rm ff}$ = 170 Myr, and that only the gas mass contained in the tidal tail ($M_{\rm gas}$ ~ 4.6 × 10$^{10}$ M$_{\odot}$, assuming $T_{\rm dust}$ = 50 K, see Table S1) is funneled to the center of W2246−0526 while most of the gas within the central few kpc remains turbulent and buoyant as suggested by the kinematics of the [C II] and CO(2→1) lines (*16*; and Figure S3), we calculate that the average accretion rate could reach $dM_{\rm gas}/dt$ ~ 270 M$_{\odot}$ yr$^{-1}$. If the self-gravitating companion galaxies are also accreted (combined $M_{\rm gas}$ ~ 4.8 × 10$^{10}$ M$_{\odot}$), then $dM_{\rm gas}/dt$ ~ 550 M$_{\odot}$ yr$^{-1}$. The free-fall time is a lower limit to the true time-scale of accretion, but we have chosen the most conservative values for the estimations of the gas masses and the free-fall time itself. That is, these have been derived using an $\alpha_{CO,ULIRG}$ conversion factor and the upper bound of $T_{\rm dust}$ ranges. Instead, if an $\alpha_{CO,MW}$ is used to calculate $t_{\rm ff}$, then $dM_{\rm gas}/dt$ can reach up to ~ 900 M$_{\odot}$ yr$^{-1}$, or even higher if the gas masses of the tidal tail and companion galaxies are obtained using the lower end of the dust temperature range.

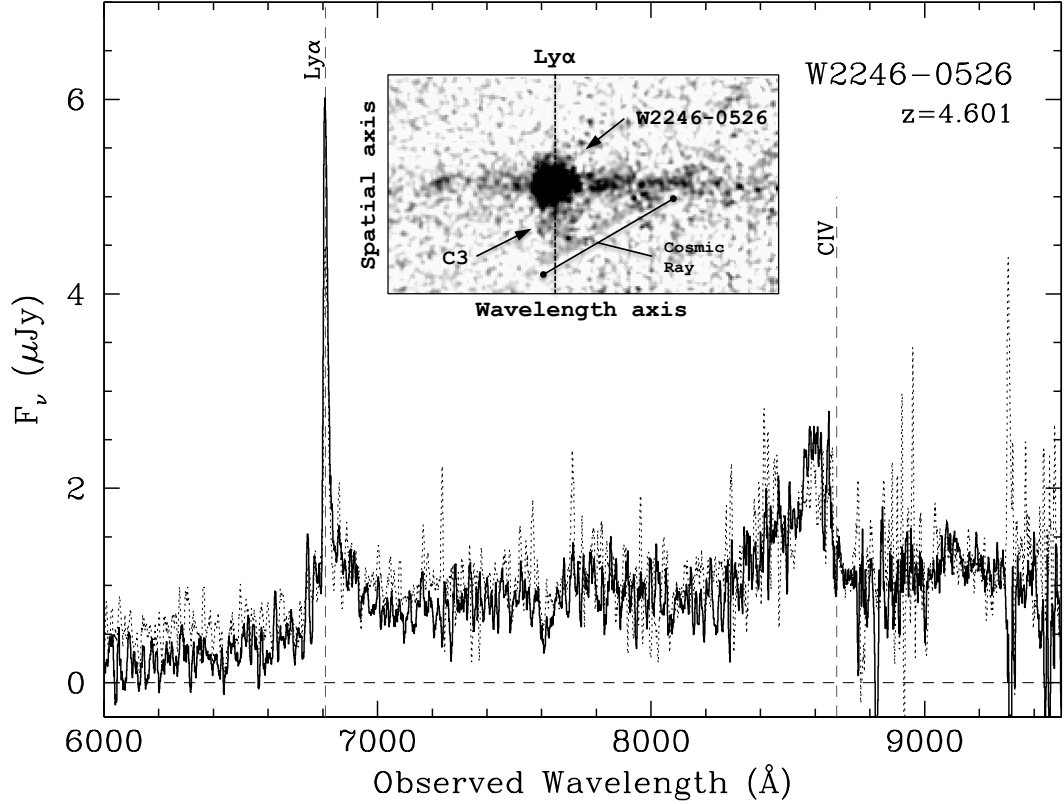

Figure S1: **Optical spectra of W2246−0526 obtained with Keck/LRIS.** $F_\nu$ is the flux density per unit frequency, which is shown as a function of the observed wavelength. The spectrum obtained in October 2013 is shown as a solid line, and the November 2010 spectrum is over-plotted as a dashed line. The inset figure shows the two-dimensional spectrum obtained in December 2015, smoothed to highlight the Lyα emission from the companion galaxy C3. The vertical axis corresponds to the slit spatial position and extends ~ 20″. The horizontal, wavelength axis is centered around 6800Å. The narrow Lyα emission in the one-dimensional spectra of W2246−0526 is asymmetric, which is typical at high redshift, with a sharp edge to the blue side of the line due to foreground and associated absorption. The indicated ALMA-derived [C II] 158 μm redshift matches the redshift derived from the peak of the Lyα line. The strongly asymmetric C IV line is much broader and is centered bue-ward of the Lyα and ALMA [C II] redshifts.

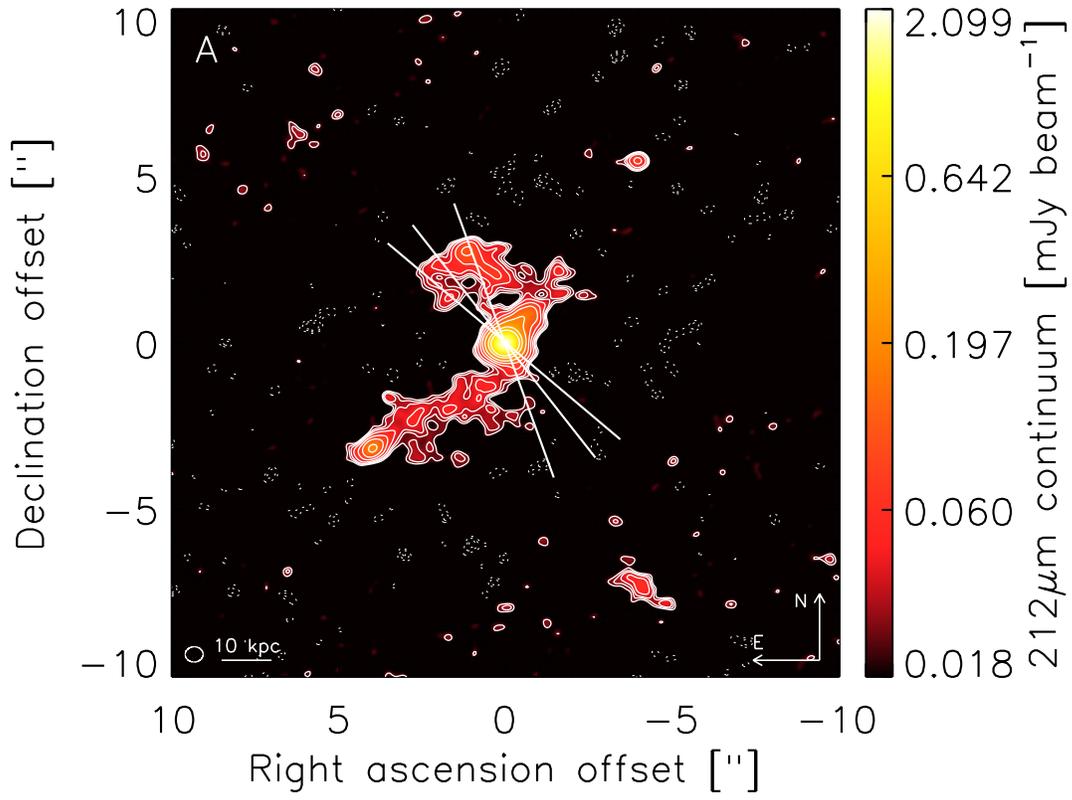
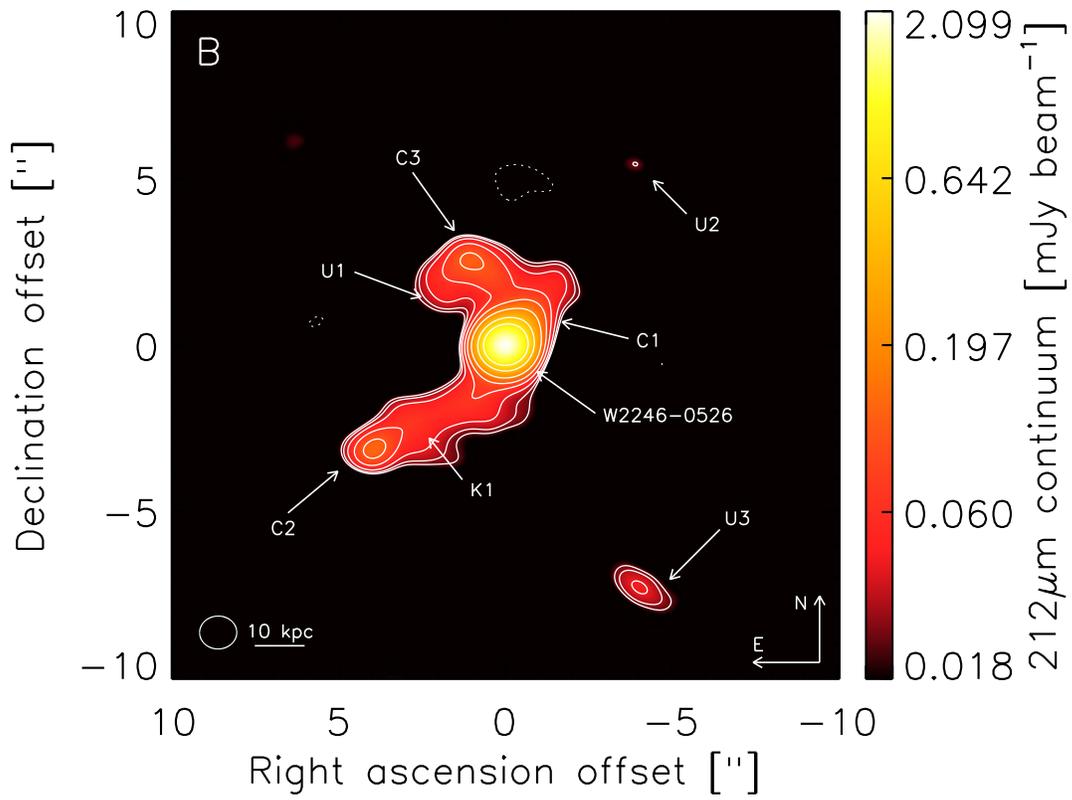

Figure S2: **ALMA 212 μm dust continuum emission maps of the W2246−0526 merger system.** (A) Same as Figure 1 but including contour levels at [+2, -2] × σ, where σ is the measured r.m.s. of the background. In addition, solid straight lines indicate the position angle of the slits used to obtain the optical spectra shown in Figure S1: 20°, 38° and 49.9° (east of north). There are more regions with negative flux than at -2.5 × σ, but most of the positive 2 × σ emission is recovered around the tidal tail and dusty bridges connecting W2246−0526 with the companion galaxies. (B) Same as Figure 1 but with a *uv*-tapering of 1″ applied to the data. The tidal tail is detected at larger angular scales. These two maps confirm the extended emission detected in Figure 1.

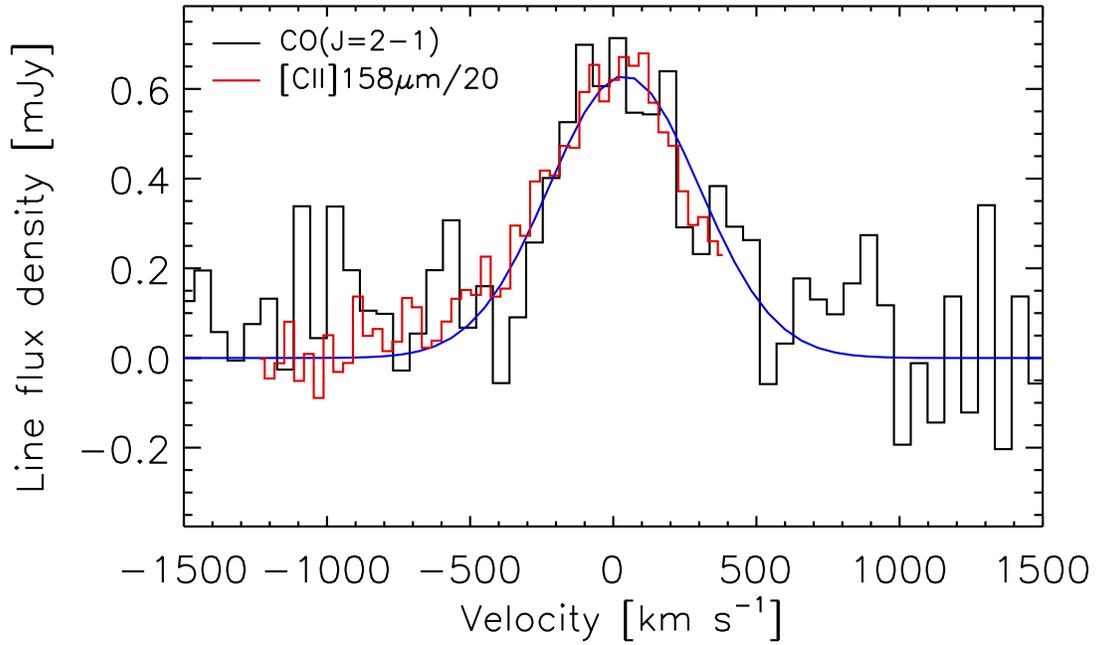

Figure S3. **The W2246−0526 spectra of the CO(2→1) (black) and [C II] 158 μm (red) emission lines.** The spectra were extracted using circular apertures of $r$ = 2.2″ (similar to the beam size of the VLA observations) and $r$ = 0.35″ (similar to the beam size of the ALMA observations), respectively, centered on W2246−0526. The [C II] spectrum has been scaled down in flux density by a factor of 20. The blue line is a Gaussian model fitted to the CO(2→1) spectrum, which results in a FWHM = 615 km s$^{-1}$. The FWHM of the [C II] line is ~ 550 km s$^{-1}$ (*16*) within the central few kpc of W2246−0526. The cold molecular gas traced by the CO(2→1) line is as turbulent as the neutral gas traced by [C II], and may be also affected by the strong feedback from the central AGN.

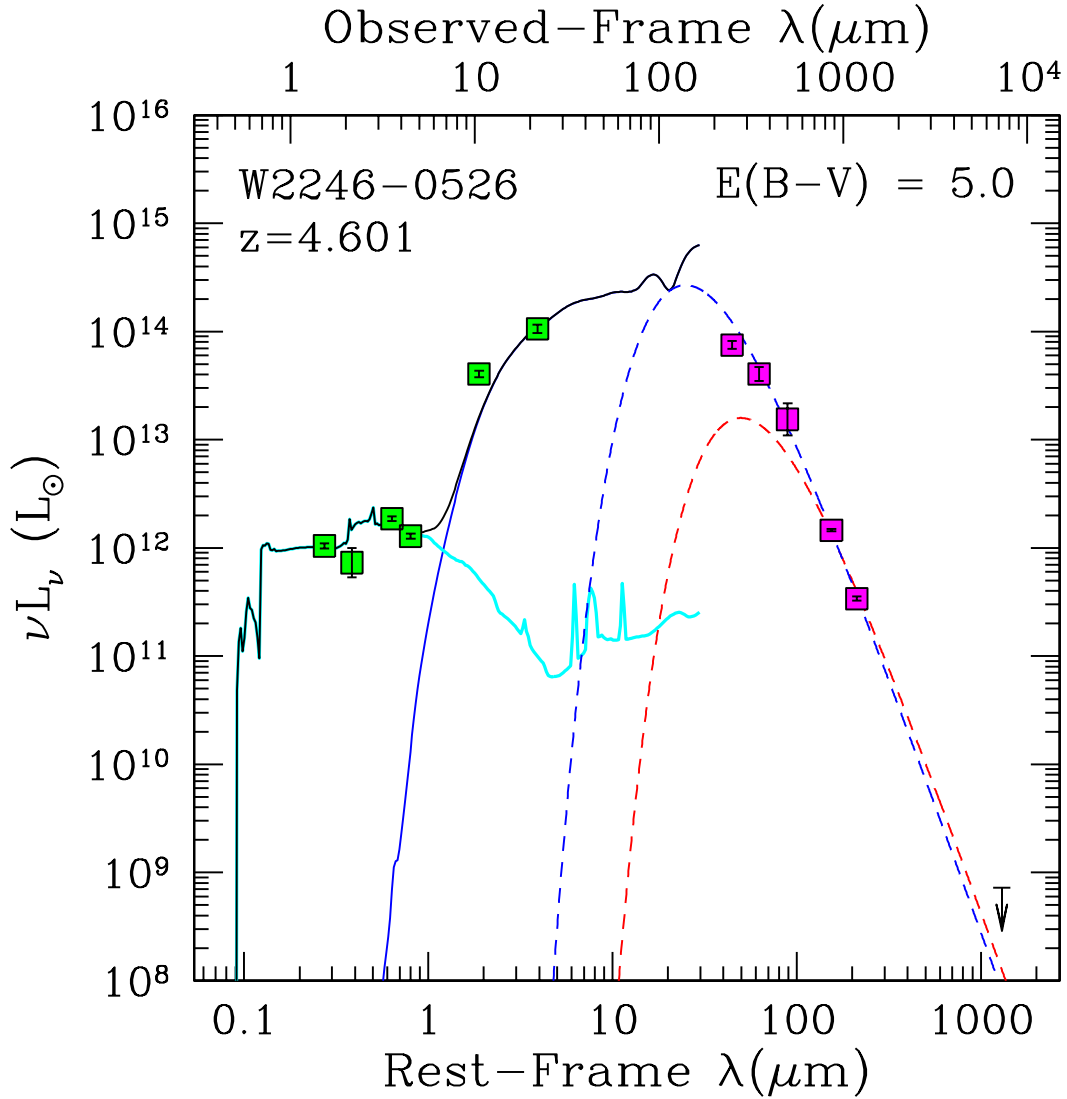

Figure S4. **Optical through far-IR SED of W2246−0526.** The mono-chromatic luminosity ($\nu L_\nu$, where $L_\nu$ is the luminosity per unit frequency) is plotted as a function of the rest-frame wavelength. The green squares show, from left to right, the observed luminosities in the F160W HST (*16*), Ks (*10*), Spitzer 3.6 and 4.5μm (*49*), and WISE W3 and W4 broad-band filters (*3*). The black line shows the best-fitting SED using the templates and algorithm of (*47*). The best fitting SED model is composed of an AGN template with reddening E(B−V) = 5.0 (dark blue solid line) and the Im galaxy template (cyan line). The magenta squares show, from left to right, the luminosities of SPIRE at 250, 350 and 500 μm (*14*), ALMA at 865 μm and 1.19 mm, and the 3σ upper limit from the VLA at 7.28 mm. Two fits to the ALMA data-points are also shown, extrapolated to shorter and longer wavelengths, using an optically-thin modified black body with an emissivity index β = 1.8 and dust temperatures of 50 and 100 K (dashed red and blue lines respectively). These correspond to the two limiting temperatures used to calculate the dust mass in W2246−0526.

Table S1. **Summary of the dust and gas mass estimates for each source in the W2246−0526 system, assuming different dust temperatures.** Column 1: Source over which the photometry was performed. Column 2: Radius of the circular aperture used to obtain the flux densities. Column 3: Observed flux density of the dust continuum at $\lambda_{rest} \sim$ 212 μm. Columns 4–6: Dust masses estimated using three different temperatures ($T_{dust}$ = 100, 50 and 25 K), assuming optically thin modified black body emission with an emissivity $\beta = 1.8$, and a dust opacity coefficient $\kappa_{\nu,dust}$ (850 μm) = 0.0484 m$^2$ kg$^{-1}$. Assuming that the dust and gas are well mixed, a standard gas-to-dust ratio typical of local, solar-metallicity galaxies (*18*) of $\delta_{GDR}$ = 100 can be applied (multiplied) to obtain the gas mass of each component. Column 7: Gas mass estimates based on the CO(2→1) emission line assuming the gas is thermalized up to the J = 2–1 transition and using a $M_{gas}$-$L_{CO(1\rightarrow 0)}$ ratio typical of local ULIRGS $\alpha_{CO,ULIRG}$ = 0.8 M$_\odot$ [K km s$^{-1}$ pc$^2$]$^{-1}$. If a ratio typical of normal star-forming galaxies were used, all values in Column 7 would be multiplied by ~5–6. The radius of the aperture used to calculate the total gas mass based on the CO(2→1) line is $r$ = 5.5″.

| Source | Aper. radius [″] | $f_\nu$ (212μm) [mJy] | $M_{dust}$ (T=100K) [M$_\odot$] | $M_{dust}$ (T=50K) [M$_\odot$] | $M_{dust}$ (T=25K) [M$_\odot$] | $M_{gas}$ CO(2-1) $\alpha_{CO,ULIRG}$ |
|---|---|---|---|---|---|---|
| W2246 | 0.5 | 2.3 ± 0.1 | 5.6 – 17 × 10$^8$ | – | – | – |
| C1 | 0.5 | 0.27 ± 0.02 | – | 2.0 – 9.6 × 10$^8$ | | – |
| C2 | 0.5 | 0.23 ± 0.02 | – | 1.7 – 8.2 × 10$^8$ | | – |
| C3 | 0.5 | 0.16 ± 0.02 | – | 1.2 – 5.7 × 10$^8$ | | – |
| K1 | 0.5 | 0.12 ± 0.02 | – | 0.9 – 4.3 × 10$^8$ | | – |
| Tail | 1.6 | 0.63 ± 0.05 | – | 4.6 – 22 × 10$^8$ | | – |
| CO beam | 2.2 | 3.9 ± 0.4 | 9.5 – 28 × 10$^8$ | | | 0.7 × 10$^{11}$ M$_\odot$ |
| Total | 4.5 | 4.9 ± 0.6 | 12 – 36 × 10$^8$ | | | 1.5 × 10$^{11}$ M$_\odot$ |